# Remote estimation of geologic composition using interferometric synthetic-aperture radar in California's Central Valley

**Kyongsik Yun, Kyra Adams, John Reager, Zhen Liu,
Caitlyn Chavez, Michael Turmon, Thomas Lu**
NASA Jet Propulsion Laboratory, California Institute of Technology
4800 Oak Grove Drive, Pasadena, CA 91109
Kyongsik.yun@jpl.nasa.gov

## Abstract

California's Central Valley is the national agricultural center, producing 1/4 of the nation's food. However, land in the Central Valley is sinking at a rapid rate (as much as 20 cm per year) due to continued groundwater pumping. Land subsidence has a significant impact on infrastructure resilience and groundwater sustainability. In this study, we aim to identify specific regions with different temporal dynamics of land displacement and find relationships with underlying geological composition. Then, we aim to remotely estimate geologic composition using interferometric synthetic aperture radar (InSAR)-based land deformation temporal changes using machine learning techniques. We identified regions with different temporal characteristics of land displacement in that some areas (e.g., Helm) with coarser grain geologic compositions exhibited potentially reversible land deformation (elastic land compaction). We found a significant correlation between InSAR-based land deformation and geologic composition using random forest and deep neural network regression models. We also achieved significant accuracy with 1/4 sparse sampling to reduce any spatial correlations among data, suggesting that the model has the potential to be generalized to other regions for indirect estimation of geologic composition. Our results indicate that geologic composition can be estimated using InSAR-based land deformation data. In-situ measurements of geologic composition can be expensive and time consuming and may be impractical in some areas. The generalizability of the model sheds light on high spatial resolution geologic composition estimation utilizing existing measurements.

## 1 Introduction

The Central Valley aquifer system is home to 6 million residents, 250 different crops, and a $17 billion per annum agricultural industry. Groundwater from the Central Valley is a valuable resource that complements surface water, especially during times of drought or limited surface water availability. It is a heterogeneous aquifer system with confined, semi-confined, and unconfined aquifers where fresh groundwater occurs in alluvial deposits down to 3000ft [1]–[3].

Due to extensive agricultural activity and land use, the groundwater system in Central Valley has steadily suffered groundwater loss, estimated to be around 125 million acre-feet of groundwater drained between 1920-2013. This groundwater extraction has led parts of the aquifer to land subsidence, as rearranging of groundwater-suspended sediment grains compacts aquifer layers. Inelastic subsidence causes severe damage within the aquifer system, such as infrastructural damage and loss of groundwater storage space.

As the groundwater is pumped, the pore space between fine-grained silts and clay decreases, lowering the land surface. Compacted land cannot store as much groundwater as it used to. Some land compactions are reversible, and some are not, depending on the geologic composition. The types of subsidence may also change over time. Once reversible subsidence regions would experience continued compaction and become irreversible regions [4]–[7]. It is important to identify the type of subsidence, because we can take appropriate actions to each type of subsidence [8]. We will be able to recharge groundwater for reversible or elastic regions, and strictly control groundwater usage for irreversible or inelastic regions.

In-situ geologic composition quantification is important in managing groundwater and preventing/recovering land subsidence, yet costly and labor intensive. Moreover, understanding geologic composition is essential for hydrological models for future predictive models of land subsidence and groundwater levels, especially in regions where extensive geologic data are not available. In this study, we aim to indirectly quantify geologic composition based on temporally changing remote land deformation information using InSAR.

## 2 Methods
### 2.0 InSAR data processing

We processed Sentinel-1 (S-1) satellite data of track 42 and 144 from 2015/03/01 to 2020/08/31 and 2014/11/08-2019/01/22, respectively to estimate ground movement associated with groundwater withdrawal/recharging in the Central Valley, California. The two tracks cover most of the central and southern Central Valley including San Joaquin Valley and Tulare Basin. The S-1 satellite constellation has been acquiring interferometric wide-swath mode data over the Central Valley with a regular interval of 12 days and a revisit time as short as 6 days using terrain observation by progressive scan (TOPS) technique. We use the JPL/Caltech ISCE software to generate the S-1 interferometry and limit our interferometric pairs to the ones with temporal baseline no more than 24 days. This mitigates temporal decorrelation. We then use a sentinel-1 stack processor to co-register all SAR single look complex (SLC) images to the reference geometry and employ the enhanced spectral diversity technique to estimate azimuth misregistration between SLC images in a stack sense. Each interferogram is corrected for topographic phase and then unwrapped and geocoded using SRTM DEM model. After generating hundreds of unwrapped interferograms for each track, we use a variant of the Small Baseline Subset InSAR time series inversion approach to solve for line-of-sight (LOS) displacement time series and mean velocity. The approach also estimates the DEM error and uses spatiotemporal filtering to suppress high-frequency troposphere noise. For more details about the InSAR processing and time series analysis, please refer to Liu et al. (2019) [9]. The cumulative LOS displacements at each image date are outputted to a GMT grid, which is resampled to a ground posting of ~2km x 2km.

### 2.1 Geologic composition

The data for geologic composition was obtained from the geotexture model published as part of the USGS Central Valley Hydrologic Model (CVHM). The geotexture model was created based on lithologic data from 8,500 borehole logs, with depth reaching down to 3000ft from land surface. The lithologic data was classified in binary bins of fine-grained or coarse-grained. This was based on the original description in the log, with coarse-grained sediments encompassing sand, gravel, pebbles, and boulders and fine-grained sediments encompassing clay, lime, loam, mud, or silt. Then the percentage of fine-grained and coarse-grained sediments were calculated in 50ft segments. The texture model arrays are organized in the same dimensions as the groundwater flow model, with 1-mi by 1-mi grid cells and 10 modeling layers based on horizontal geologic characteristics. Roughly 20,000 model cells are active within the Central Valley. Readers are pointed to Faunt (2009) [10] for more details on the geotexture model.

Coarse grain percent of each model layer from 1 to 10 in northern and southern Central Valley regions was used as the prediction target (Supplementary Figure 1). Coarse-grained soil is defined as containing no more than 50% fine grains (i.e., silt and clay, or particles smaller than 0.075 mm).

### 2.2 Data interpolation

We first integrated multimodal data including InSAR, groundwater, precipitation, and geologic composition by interpolating data with the same spatial and temporal resolutions (every 2 weeks on a 2kmX2km grid) (Supplementary Figures 2 & 3). We then identified regions with different temporal dynamics of land displacement, groundwater depth, precipitation, and geologic composition (Supplementary Figure 4). Some areas (e.g., Helm) with coarser grain geologic compositions exhibited potentially reversible land transformations (elastic land compaction).

## 2.3 Geologic estimation models

We used long short-term memory (LSTM) as a recurrent network component. Conventional recurrent neural networks still have significant practical problems caused by exponential decay of gradient descent, which hinders learning of long-term relationships between time points. LSTM is a special type of recurrent neural network that can learn long-term dependencies through selective memory consolidation [11]. We used 3 convolutional input layers, 6 recurrent layers, 1 fully connected layer, and 1 softmax layer [12], [13]. Model training aims to minimize the error function E, the mean squared error (MSE) that quantifies the difference between the estimated (neural network outputs) and the true 10-layer geologic compositions (ground truth in-situ data). Input data include InSAR subsidence data, covering 8818 different locations, and 132 biweekly time points (5 years). Here, the coarse grain percent of 10 layers was estimated.

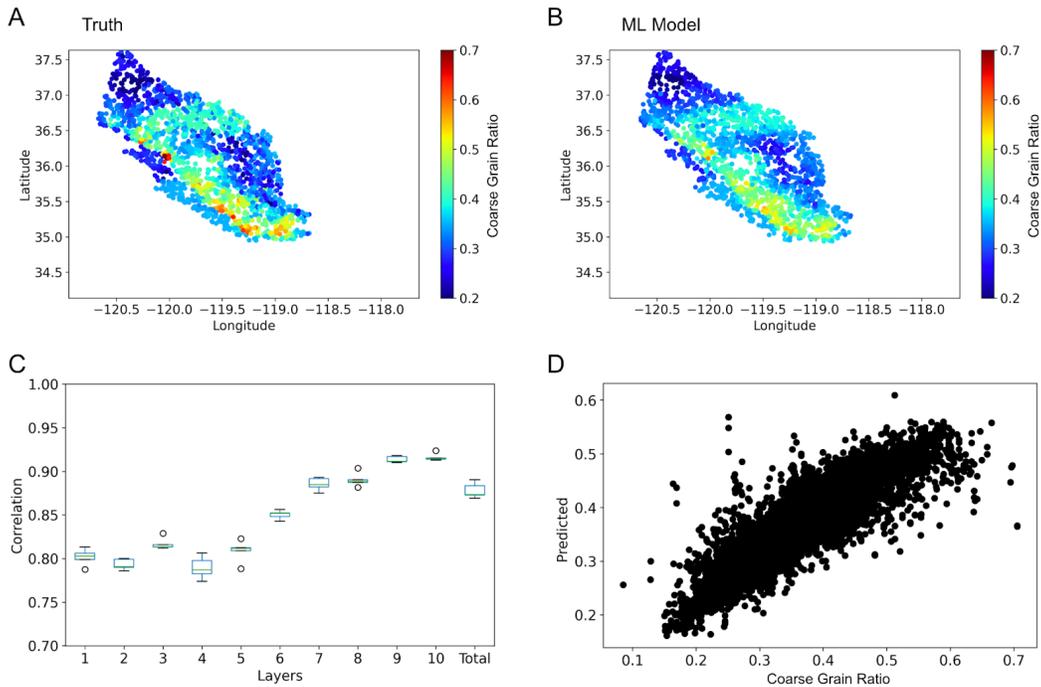

Figure 1. Geologic composition prediction using InSAR land deformation data. (A) Ground truth coarse grain ratio of the entire layer and (B) estimated coarse grain ratio. (C) Correlation between model output and ground truth at different layers of geologic composition. (D) Scatter plot between the ground truth and estimated geologic composition of the entire layer (R=0.88).

## 3 Results

We found that the InSAR remote sensing data had predictive power for geologic composition using deep neural networks (correlation coefficient R=0.88) (Figure 1). A decision tree (R=0.65) and random forest model (R=0.85) was tested as baseline. We also achieved significant accuracy with only 40% of the training data (R=0.80), suggesting that the model can be generalized to other regions for indirect estimation of geologic composition. We performed an estimation with distant data sampling (minimum distance between samples was 10km) to reduce the impact of spatial correlation of adjacent data points, and we found a slightly degraded performance (R=0.83) (Supplementary

Figure 5).

## 4  Discussion

In this study, we showed that geological composition can be estimated remotely using InSAR land deformation data. In-situ measurements of geological composition are critical to understanding hydrology and monitoring groundwater availability. However, in-situ measurements are expensive and time consuming. If geologic composition can be measured remotely using this model, high spatial resolution geologic composition can be quickly quantified only with InSAR satellites without in-situ measurements. The next step is to apply this model to other regions, including US High Plains and North China Plains, to evaluate its generalizability [14].

As a further analysis to determine which time of year contributed the most to the estimation of geologic composition, we performed a leave-one-month-out 10-fold cross-validation performance test (Figure 2). When we excluded the October and December data from the estimation model, we found a significant decline in correlation, indicating that this month contributed the most to estimating the geological composition. Most of the precipitation occurs in late autumn and winter, and precipitation has influenced time-series changes in InSAR land deformation, indirectly indicating the inner geological composition of the Central Valley.

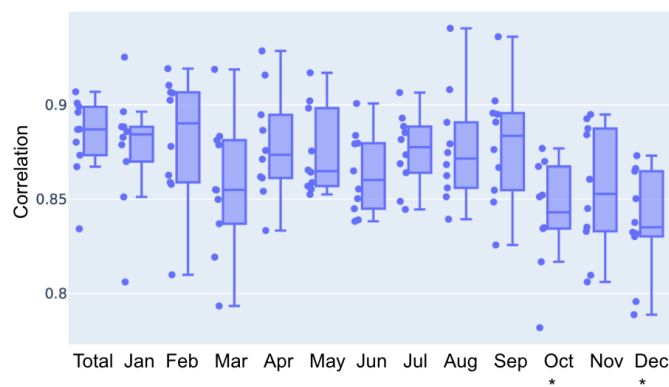

Figure 2. Leave-one-month-out 10-fold cross-validation performance test. Correlation degradation was computed as each month was excluded from the input parameters of the model. October ($p=0.00175$) and December ($p=0.000654$) showed significant degradation compared to total data correlation. The statistical significance cut-off is at $p=0.004$ considering Bonferroni multiple comparisons (0.05/12 comparisons=0.004).

One caveat of this study is the lack of ground truth geologic composition data for independent model validation. In-situ geologic composition data from other regions (e.g., US High Plains, North China Plains) will be required for further testing. At the same time, the lack of geologic composition data points to the advantage of a significant potential applicability of this model. Our suggested model can be applicable to future missions such as NISAR and NASA's Decadal Survey Designated Observables like Mass Change (MC) and Surface Deformation and Change (SDC) as an indirect geologic composition product [15].

## Acknowledgments

The research was carried out at the Jet Propulsion Laboratory, California Institute of Technology, under a contract with the National Aeronautics and Space Administration.

# Supplementary Information

Supplementary Figure 1. Coarse Grain (%) of each layer from 1 to 10 in northern and southern Central Valley regions.

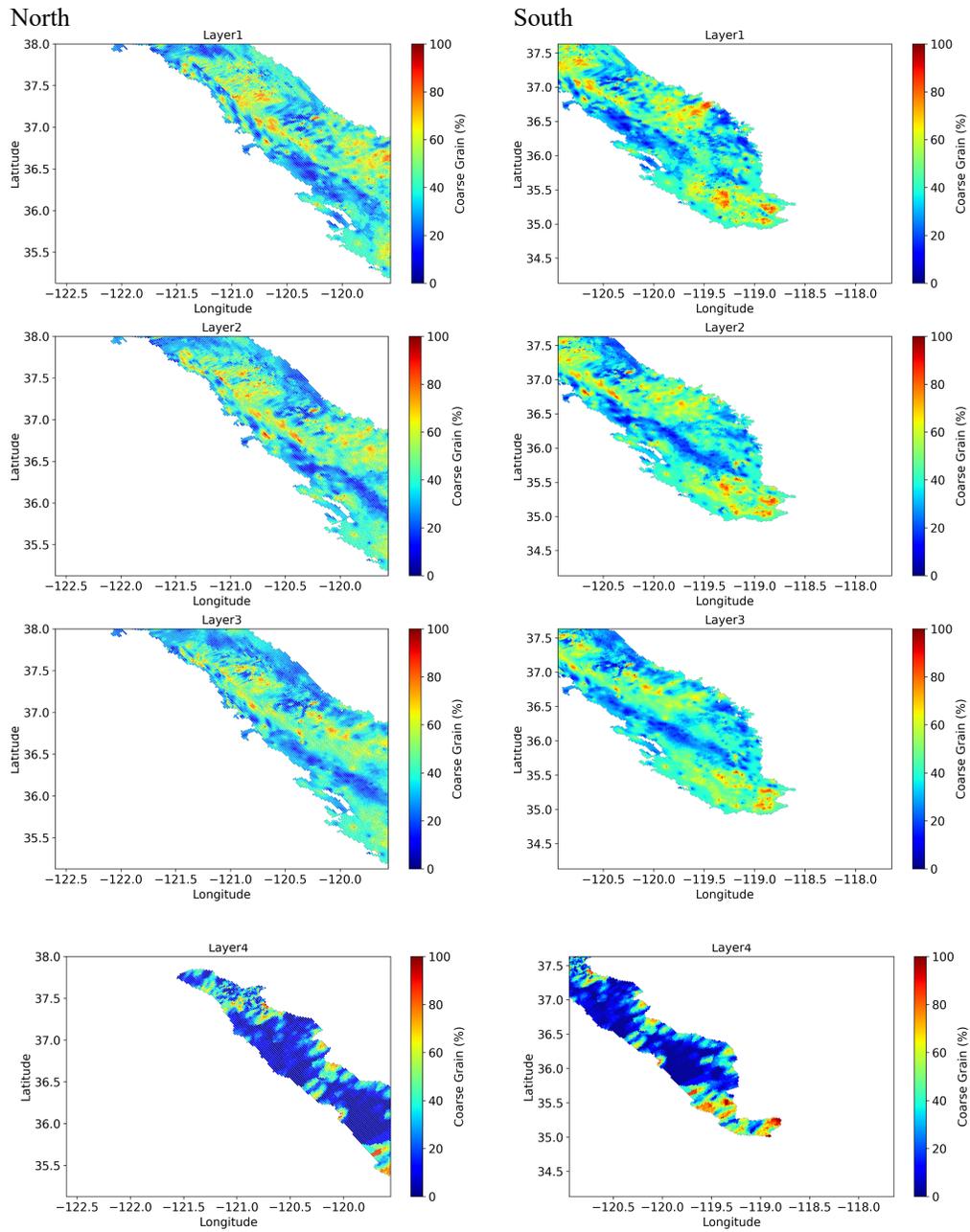

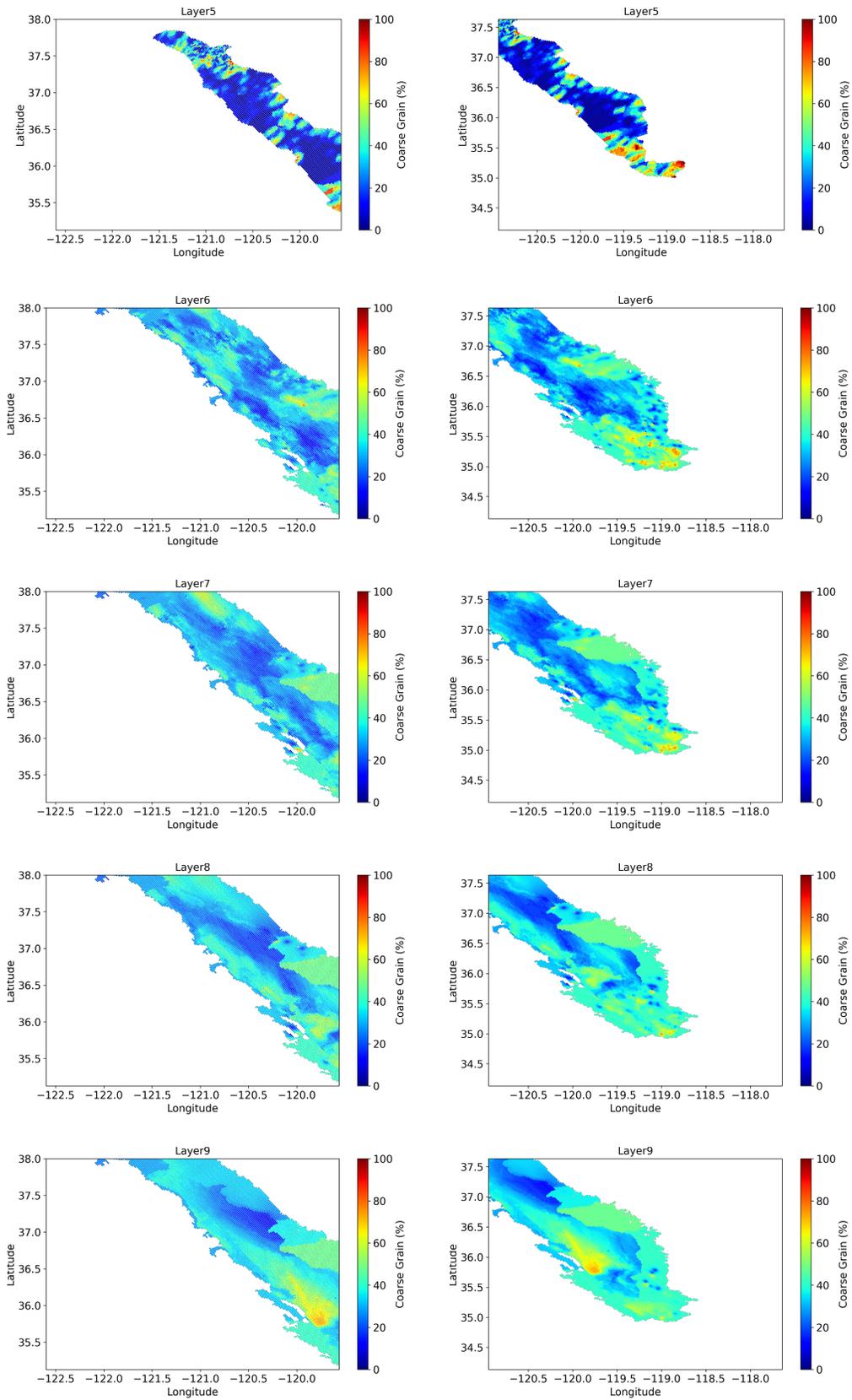

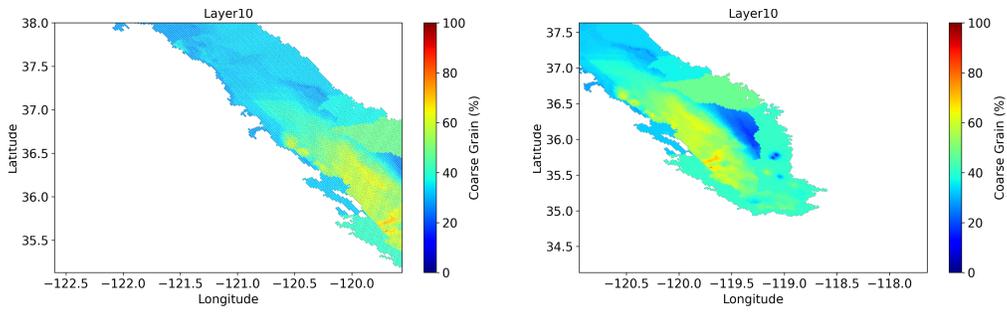

Supplementary Figure 2. Northern Central Valley InSAR land displacement (March 1, 2015 ~ August 31, 2020)

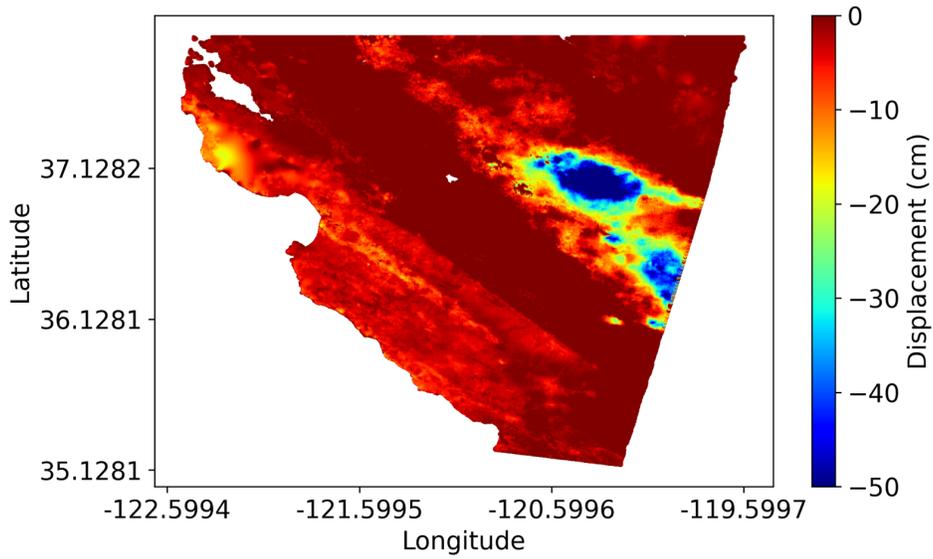

Supplementary Figure 3. Southern Central Valley InSAR land displacement (November 8, 2014 ~ January 22, 2019)

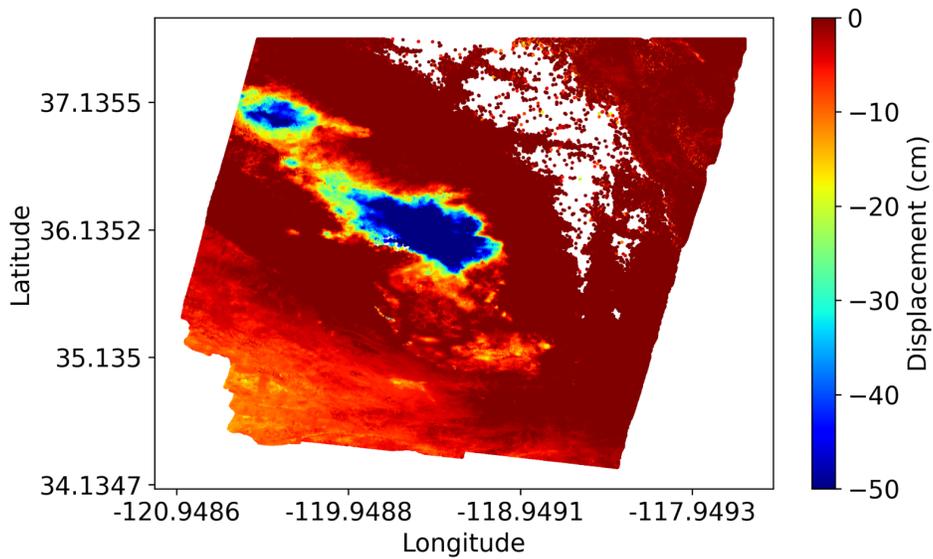

Supplementary Figure 4. Two representative regions of the Central Valley with significant subsidence with different characteristics. (A) Chowchilla has been shown to maintain monotonically decreasing land displacements, less fluctuating groundwater depth, relatively low precipitation, and high fine-grain ratio across the middle soil layers (Displacement = -22.47 ± 10.66, Groundwater (ft) = 95.53 ± 27.69, Rain (mm) = 0.84 ± 0.80, Coarse Grain (%) = 27.44 ± 10.13). (B) Helm, on the other hand, exhibited fluctuating land displacements, relatively large seasonal changes in groundwater depth, high precipitation, and a higher overall coarse-grain ratio across all soil layers (Displacement = -40.95 ± 14.49, Groundwater (ft) =160.52 ± 65.82, Rain (mm) = 3.48 ± 3.15, Coarse Grain (%) = 40.01 ± 1.67).

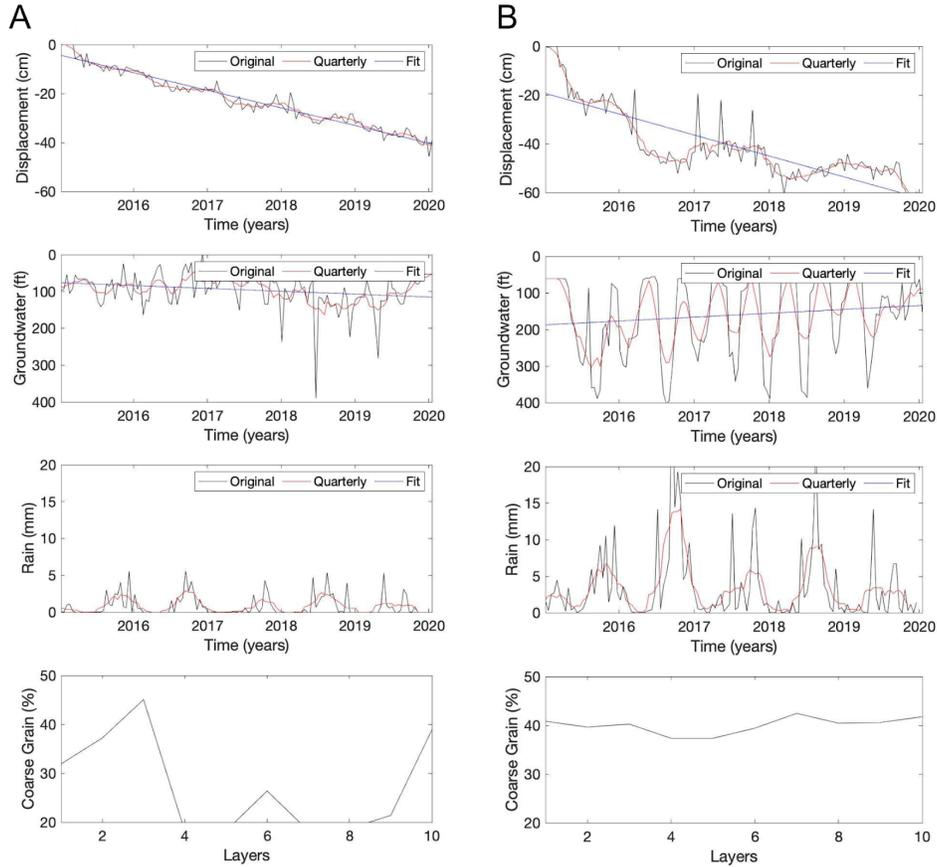

Supplementary Figure 5. Geologic composition prediction with distant data sampling (minimum distance between samples was 10km) using InSAR land deformation data. Distant data sampling was performed to reduce the impact of spatial correlation of adjacent data points. Total prediction performance dropped from 0.88 to 0.83, but remained largely unchanged.

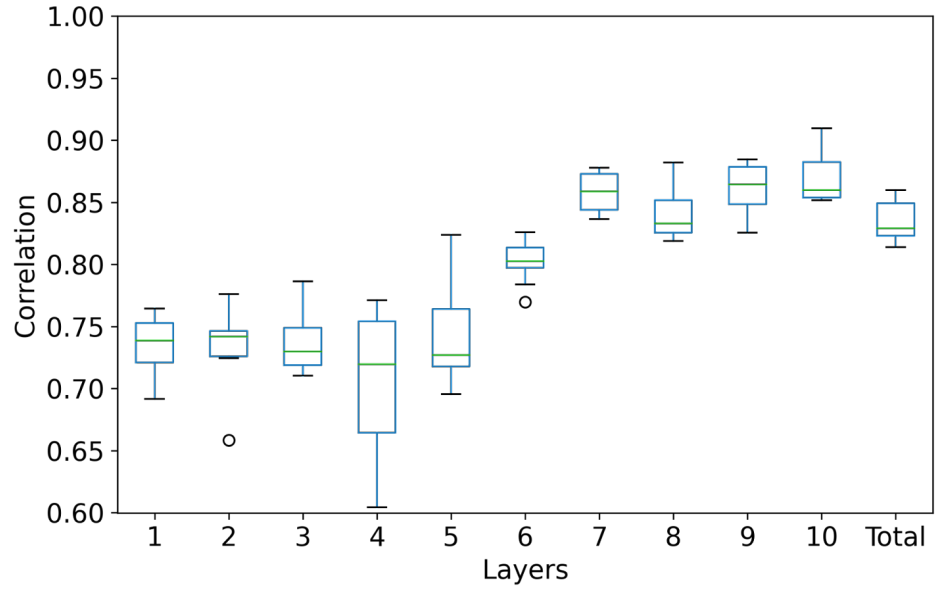